\documentclass[aps,floatfix,twocolumn,final,prb,showpacs]{revtex4}

\usepackage{graphicx}

\begin{document}

\title{Spin order in the one-dimensional Kondo and Hund lattices}

 \author{D.J.$\;$Garcia}
\affiliation{Instituto de Fisica Gleb Wataghin, Unicamp, 
Campinas (6165), S\~ao Paulo, Brasil}
\affiliation{Instituto Balseiro and Centro At\'{o}mico Bariloche, CNEA, 8400 San Carlos 
de Bariloche, Argentina.} 
 \author{K. Hallberg and B. Alascio}
\affiliation{Instituto Balseiro and Centro At\'{o}mico Bariloche, CNEA, 8400 San 
Carlos de Bariloche, Argentina.} 
\author{M. Avignon} 
\affiliation{Laboratoire
d'Etudes des Propri\'{e}t\'{e}s Electroniques des Solides, Associated
with Universit\'{e} Joseph Fourier, C.N.R.S., BP\ 166, 38042 Grenoble
Cedex 9, France.}

\begin{abstract} 

We study numerically the one-dimensional Kondo and Hund lattices consisting of
localized spins interacting antiferro or ferromagnetically with the itinerant
electrons, respectively. Using the Density Matrix Renormalization Group we find, for
both models and in the small coupling regime, the existence of new magnetic phases
where the local spins order forming ferromagnetic islands coupled
antiferromagnetically. Furthermore, by increasing the interaction parameter $|J|$ we
find that this order evolves toward the ferromagnetic regime through a spiral-like
phase with longer characteristic wave lengths. These results shed new light on the
zero temperature magnetic phase diagram for these models.

\end{abstract}


\maketitle

The interplay between charge and spin degrees of freedom in strongly
correlated systems has triggered enormous interest in recent years due to the rich variety of phases found in a plethora of compounds. Charge and spin
superstructures with a doping dependent wave-vector were found for example
in La$_{2-x}$Sr$_{x}$NiO$_{4}$ using neutron scattering\cite{hayden} and
electron diffraction\cite{chen}. Stripe formation together with
incommensurate spin fluctuations in High-Tc superconductors can also be
regarded as a manifestation of similar phenomena\cite{tranq} as well as the
charge and spin ordering found in many of the doped manganese perovskites.
Another large group of compounds, the heavy fermion materials, present
various types of ground states including antiferromagnetically ordered
states, the normal heavy fermion state as well as superconducting and
insulating phases. Heavy-fermion systems and Kondo insulators are typical
examples of systems in which the interactions between conduction electrons
and quantum localized spins are essential\cite{hewson,tsune97}. Their
physical properties result from an antiferromagnetic coupling $J$ between
these two types of particles, the so-called Kondo lattice model (KLM). The
corresponding Hamiltonian has the well-known form:

\begin{equation}
H=-t\sum_{\langle i,j\rangle,\sigma }c_{i\sigma }^{\dagger }c_{j\sigma
}-J\sum_{i}\vec{S}_{i}\cdot\vec{\sigma }_{i}
\end{equation}

The first term represents the conduction electron hopping between
nearest-neighbor sites, $c_{i\sigma }^{\dagger }$( $c_{i\sigma }$) being
standard creation (annihilation) operators.\ In the second term the exchange
interaction $J$ is antiferromagnetic ($J<0$), and $\vec{\sigma }_{i}=\frac{1}{2}\sum_{\sigma ,\sigma ^{\prime }}c_{i\sigma }^{\dagger
}\vec{\tau}_{\sigma \sigma ^{\prime }}c_{i\sigma ^{\prime }}$, ($\vec{\tau} _{\sigma
\sigma^{\prime }}$ are Pauli matrices).

It is interesting to note that, in recent years, the same model Hamiltonian
with ferromagnetic coupling ($J>0$) has been considered to contain the basic
physics of manganites exhibiting the ``colossal" magnetoresistance effect\cite
{coey99,tokura,manganites,dagotto01}. In this case, both localized spins and
itinerant electrons originate from manganese $d$-states.\ The system is
assumed to contain essentially Mn$^{4+}$ ions with three localized $t_{2g}$
orbitals represented as local spins $\vec{S}_{i}$ and additional
itinerant electrons in the $e_{g}$ orbital. Due to the strong Hund coupling
the spin of the $e_{g}$ electron is constrained to be parallel to the local
spin on that site. Hund's rule together with the hopping term give rise to
the ``double-exchange" (DE) interaction that favors ferromagnetic ordering of
the local spins\cite{Zener}. In recent literature this model is often referred to as 
the
ferromagnetic Kondo lattice (FKLM), however to avoid confusion with the
Kondo model, we will call it Hund model (HM).
The DE mechanism only requires that the system is away from half filling and is independent of the sign of $J$\cite{Anderson}.

We will study the Hamiltonian (1) for both, antiferro and ferromagnetic 
couplings, considering $S=1/2$ localized quantum spins. To this end
we will use the density-matrix renormalization group (DMRG)\cite{nos} with
open boundary conditions for chains of different sizes. 
We implemented the finite version of the DMRG algorithm
reaching chains sizes of 36 sites (the discarded weight being less
than $10^{-5}$). The different phases are characterized through the local spin-spin 
correlation functions and its Fourier transform, the following spin structure factor:
\[
S(q)=\frac{1}{L}\sum_{i,j}e^{iq(R_{j}-R_{i})}\left\langle 
\vec{S}_{i}\cdot\vec{S}_{j}\right\rangle 
\]
where $L$ is the number of sites in the system and $n$ the number of conduction electrons per site.

In Fig.1 we present the phase diagram of the one-dimensional Kondo and Hund models 
which we
propose from our numerical results for several commensurate fillings (marked with 
dashed lines), improving previous results \cite
{tsune97,yunoki98,dagotto98,caprara,riera,honner,shibata}.  As can be seen, except for
a scale factor, both phase diagrams present great similarities.  The half-filled case
$n=1$ is pathological in both models whose ground state is very different from the
$n\neq 1$ case: characterized by a spin gap, however with a different behavior as a
function of $|J|$ in the Kondo and Hund cases, it has been referred to as a
``spin-liquid phase"\cite{tsune2}.  For the Hund model it scales to the $S=1$ chain with 
a
Haldane gap.  The transition to the FM phase in the Kondo model coincides with the 
previous
work mentioned.  However, for the Hund model we find that the border lies at slightly
larger values of $J$ as compared with Ref. \onlinecite{dagotto98} due to the larger
system sizes considered here. We also include the phase separated regime
(PS)\cite{dagotto98}.

At low $|J|$ a phase qualified as ``paramagnetic'' in the KLM and ``incommensurate'' in
the HM had been identified with exact diagonalization and
DMRG\cite{yunoki98,dagotto98,caprara,moukouri,xavierPRB,shibata}, this phase is however
much less understood than the ferromagnetic phase. In these references the local
spin-spin correlations are calculated and $S\left(q\right) $ shows a peak at the wave
number corresponding to $2k_{F}$ of the conduction electrons.  However, no scaling was
performed and the existence of what we call ``island phases" (see below) was missed.
Recently, the existence of a ``spin dimerized'' phase has been reported\cite{xavier}
for the Kondo model at quarter-filling ($n=1/2 $), through the order parameter
$D(i)=\left\langle \vec{S}_{i}\cdot\vec{S}_{i+1}\right\rangle $. The
spin structure is of the island type $...\uparrow \uparrow \downarrow \downarrow ...$
similar to the one we have identified previously for the DE-Superexchange model
\cite{garcia00}. Our present calculations reproduce the results for $D(i)$.
However, the absence of a spin gap\cite{xavier} and our results on the long distance spin-spin correlation functions suggest a non-spontaneously broken translation symmetry state compatible with the generalization of the LSM theorem\cite{affleck}.
We find that similar structures are present for
other commensurate fillings $n=1/4,1/3,2/3$ and $3/4$ as well. However, we do not find 
that the ``island''
phases stable at small $|J|$ transform directly into the FM phase\cite{xavier}.  
Instead, they evolve toward the FM regime through an intermediate spiral-like phase
with a wave vector $q$ that changes from $q=2k_F=\pi n$ (island) to $q=0$
(F) as $|J|/t$ increases.  A spiral phase has been proposed analytically in a
semi-classical model showing a similar evolution toward the FM phase\cite{fazekas}.

\begin{figure}[htbp]
\includegraphics[height=6cm,width=8cm,clip]{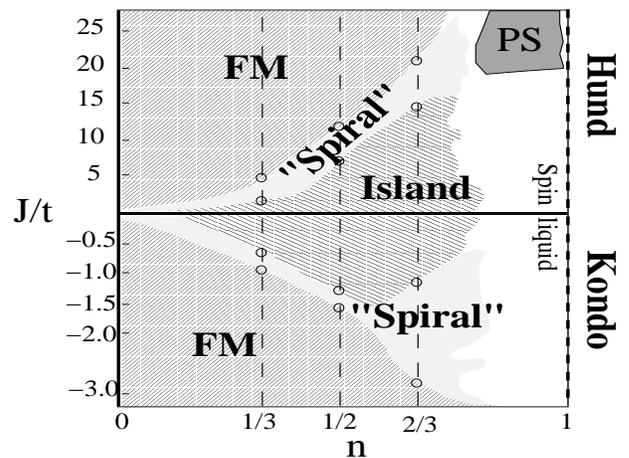}
\caption{Magnetic phase diagram for the KLM and HM. The various
phases are described in the text. The circles
indicate the crossover values. }
\label{diagfaseKH}
\end{figure}

Let us first present the quarter-filled case $n=1/2$ (Fig.2). For the Hund
model and large $J$, {\it i.e} $J/t\gtrsim 13$, the FM phase is clearly
identified with a peak of $S\left( q\right) $ at $q=0$, and the total spin
(considering only localized spins) is $S_{T}=N_{s}/2$, where $N_{s}$ is the
number of localized spins in the system. Each electron forms a triplet state 
$S=1$ with the localized spins and all localized spins are ferromagnetically
ordered. For intermediate values of $J$ ($6\lesssim J\lesssim 10$) the FM
phase gives rise to a ``spiral" phase (Fig. 3), characterized by two broad
peaks located at incommensurate values of $q$ which evolve inwards (the
momentum of the peak grows from $q=0$ to larger values and another peak at
a symmetric point with respect to $q=\pi $ moves towards smaller values).
When $J$ decreases further ($J/t\lesssim 6$) the spin structure transforms
into an ``island'' structure with a more defined peak of $S\left( q\right) $
at $q=n\pi =\pi /2$ as shown in Fig. 2(f) indicating a four-site
periodicity. This phase remains stable down to very small values of $J/t$,
it has total spin $S_{T}=0$ and zero spin gap.

Similar results are obtained in the Kondo case ($J<0$). The island-type phase is conserved for a
small Kondo coupling ($J/t\gtrsim -1.3$) and the ferromagnetic phase is recovered for $J/t\lesssim
-1.6$ (Fig. 2 (a), (c) and (e) ).  In the strong Kondo coupling limit the ferromagnetic state
differs from the Hund case in the sense that now, the conduction electrons form singlet states
($S=0$) with the localized spins, so that the system contains $N_{s}/2$ singlets and $N_{s}/2$
unpaired localized spins.  
The unpaired spins are ferromagnetically coupled and the
total spin is now $S_{T}=N_{s}(1-n)/2=N_{s}/4$.
In the ferromagnetic phase, our results for the ground-state energy compare well with those obtained with
the effective Hamiltonian in the strong coupling limit ($|J|/t\rightarrow \infty $)\cite{salerno} i.e.\ $-J/8-2t/\pi $ in the Hund case and $3J/8-t/\pi $ in the Kondo case.\ Triplet states in the Hund case conserve the hopping $t$ while singlet states in the Kondo case acquire a reduced hopping $t/2$.

The weight of the peak of the spin correlation $S(q^{*})$ for the island (and
ferromagnetic) phases increases with the chain sizes indicating a 
quasi-long-range-order in this case. Instead, for the spiral-like phase, this weight 
remains constant, indicating a short range correlation. A similar behavior for these phases 
is found for the other fillings considered.

\begin{figure}[htbp]
\includegraphics[width=8cm,clip]{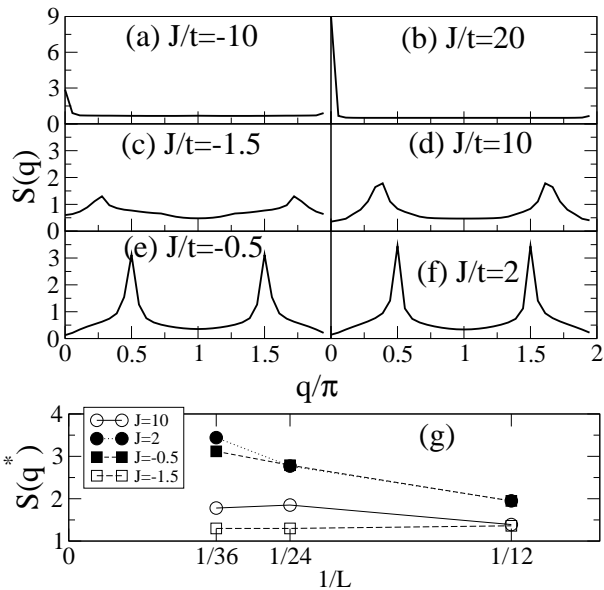}
\caption{Spin structure factor S(q) for the quarter-filled case ($n=1/2$) for the Kondo
(left panel) and Hund (right panel) models. From top to bottom: the ferromagnetic,
``spiral" and island phases, appearing as $|J|$ is decreased. The lower panel shows the
size dependence of the peak weight $S(q^{*})$ in island (full symbols) and ``spiral" 
phases (empty symbols).
\label{Sqn1s2}}
\end{figure}

In order to visualize the order in real space we show in Fig. 3 the spin-spin
correlation functions for the ``island" and ``spiral" phases. For the former case, the
correlations change sign quite abruptly every two sites indicating that the structure
is effectively of the island type $\uparrow \uparrow \downarrow \downarrow $ mentioned
above with quasi-long range ordering.  The nearest neighbor correlation $\left\langle \vec{S}(i)\cdot\vec{S}(i+1)\right\rangle $
changes abruptly from site to site, with a slight border effect. Here we reproduce the 
results of Ref. 22.
In the latter case,
instead, the ``spiral" order is clearly identified and differs qualitatively from the
island order.  The nearest neighbor correlation $\left\langle \vec{S}(i)\cdot\vec{S}(i+1)\right\rangle $ remains ferromagnetic
over the chain, with a small oscillation of period 2 suggesting a dominant spiral-like 
state with a remnant of the island order, which diminishes with $J$.

\begin{figure}[htbp]
\includegraphics[width=8cm,clip]{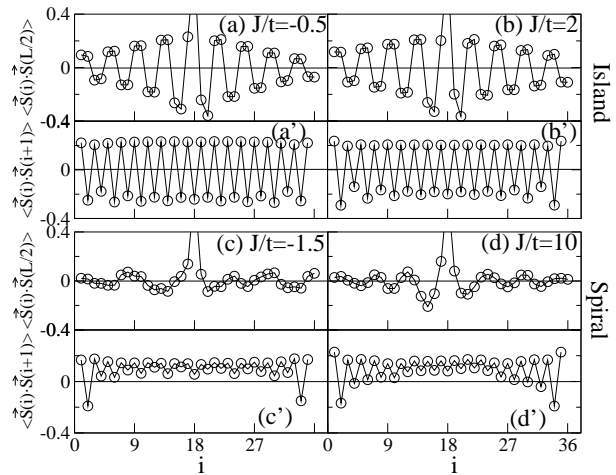}
\caption{Real-space local spin-spin correlations for $n=1/2$ and the Kondo
(left) and Hund (right) models. The qualitative difference between the
``island" (top panel) and ``spiral" (lower panel) phases can be seen.}
\end{figure}

Similar ``island'' phases are clearly evidenced for $n=1/3$ at intermediate values of
the coupling for both models ($-0.5\lesssim J/t\lesssim 1$). Typical examples are for
$J/t=1$ and $J/t=-0.3$ (see Fig. 4(c)-(d)). $S\left( q\right) $ shows a clear peak at
$q=n\pi =\pi /3$ and the spin correlation in real space presents an ``island''
structure of three ferromagnetic spins coupled antiferromagnetically between islands,
basically $\uparrow \uparrow \uparrow \downarrow \downarrow \downarrow $ (not shown).
The ferromagnetic phase as discussed above is clearly recovered for $ J/t\lesssim -1$
and $J/t\gtrsim 5$.  In the intermediate region (for values of $J$ between these
intervals) we find again the ``spiral" phase (Fig. 4 (a)-(b)).

A similar behavior occurs for $n=2/3$ as shown in Fig.4 (e)-(h).  The island structure
of the type $\uparrow \uparrow \downarrow \uparrow \uparrow \downarrow$ is clearly
identified for $-1\lesssim J/t\lesssim 10$ in $S(q)$ with a peak at $q=2\pi/3$. 
This island structure has a ferrimagnetic character\cite{noteferrimagnet} and
we suggest that this configuration corresponds to the new ``ferromagnetic" phase
reported in Ref. \onlinecite{McCulloch}. For $15\lesssim J\lesssim 21$ we observe a
``spiral" phase. 
In the Kondo model we find that the island phase
transforms into the FM regime through another intermediate phase, in the region
$-3\lesssim J\lesssim -1$, with double wave length (see below).

\begin{figure}[htbp]
\includegraphics[width=8cm,clip]{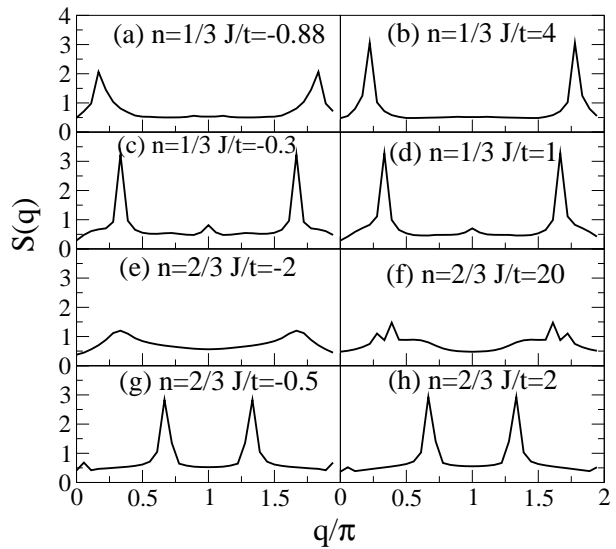}
\caption{Spin structure factors for both, Kondo (left) and Hund (right)
models showing the ``spiral" ((a)- (b) and (e)-(f)) and island ((c)-(d) and (g)-(h)) 
phases for different fillings.
\label{n1s3y2s3}}
\end{figure}

\begin{figure}[h]
\includegraphics[width=8cm,clip]{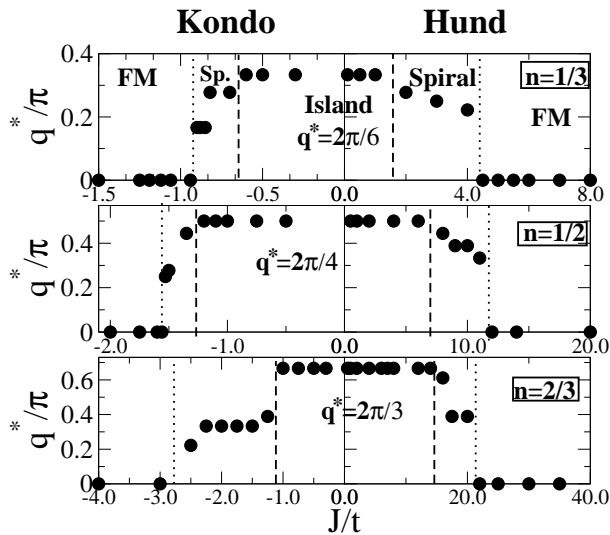}
\caption{Evolution of the momentum where the spin structure factor is
maximum as a function of the interaction strength $J$ covering both models
for different fillings.\label{Qes}}
\end{figure}

To illustrate the crossover between the different phases we plot in Fig. \ref{Qes} the behavior of the wave vector where the spin structure factor is maximum, 
$q^{\ast }$, as a function of $J$ for several fillings. 
Here it is clearly seen that for small $|J|$ the island phases show up with their characteristic wave vector $q^{\ast }=2k_{F}=n\pi $. 
This phase is stable for a certain region until the ``spiral" phase takes over for larger $|J|$.\
The wave vector of the ``spiral" phase decreases to zero as $|J|$ increases, leading finally to the 
FM phase at sufficiently large values of the interaction parameter. 
We observe that this behavior is qualitatively
similar to the ``unwinding'' of the spiral phase towards the FM order
obtained by Fazekas and M\"{u}ller-Hartmann\cite{fazekas}.
For $n=2/3$ in the Kondo case we find that $q^{\ast }$ remains fixed at $q=\pi/3$ in the region labeled as ``polaronic liquid" in Ref. \onlinecite{McCulloch}.
In this case, from the real-space spin-spin correlations it is difficult to distinguish clearly between 
a ``spiral" and an island of type $\uparrow \uparrow \uparrow\downarrow \downarrow \downarrow $ but 
with short range order.

In conclusion we have studied numerically, using the DMRG, the Kondo and Hund models for
localized spins interacting with itinerant electrons. In addition to the ferromagnetic phase 
at
large $|J|$, we find the existence of island and ``spiral" phases within the
``paramagnetic regime'' in these models. 
Both phases differ qualitatively as seen in different correlation functions. The island 
phase has quasi-long range order and zero spin gap.
Furthermore we show how the ground-state
evolves from the low $|J|/t$ ``island'' $2k_F$-phase to the FM regime through a
spiral-like phase at intermediate couplings. By carefully analyzing the finite-size 
scaling, we conclude that all phases obtained are  {\it not due to Friedel 
oscillations} of the open boundaries.
 Based on the results for commensurate
fillings we suggest the phase diagram shown in Fig.1. We would like to point out that 
the
difference in the scale of $J$ between the Kondo and Hund models has to be a
consequence of the quantum nature of the localized spins, so that a complete
understanding of the phase diagram requires a full quantum description of the 
Hamiltonian.

D.J. Garcia acknowledge discussions with E. Miranda and J.C. Xavier. We also thank C. 
Batista for fruitful comments.
We acknowledge partial support from the Collaboration Program between France and Argentina CNRS-PICS 1490, CONICET-PICT, PIP (CONICET) 437/98 and Fundaci\'{o}n Antorchas Project 14116-168. D. G. and K. H. acknowledge full and B. A.  partial support by CONICET.

\end{document}